\documentclass[twocolumn,prd,superscriptaddress,showpacs,floatfix,%
preprintnumbers,nofootinbib,amsmath,amssymb,amsfont]{revtex4-1}  

\usepackage{bm}  
\usepackage{graphicx}  
\usepackage{hyperref}

\newcommand{\ben}{\begin{equation}}
\newcommand{\een}{\end{equation}}

\newcommand{\be}{\begin{equation}}
\newcommand{\ee}{\end{equation}}  
\newcommand{\bea}{\begin{eqnarray}}
\newcommand{\eea}{\end{eqnarray}}  
\newcommand{\gag}{g_{a\gamma}}

\begin{document}

\title{Probing axion-like particles with the ultraviolet photon polarization \\ from
active galactic nuclei in radio galaxies}

\author{Dieter Horns}
\email{dieter.horns@desy.de}
\affiliation{II Institut f\"ur Experimentalphysik, Universit\"at Hamburg,
Luruper Chaussee 149, 22761 Hamburg, Germany} 

\author{Luca Maccione} 
\email{luca.maccione@lmu.de}
\affiliation{Arnold Sommerfeld Center, Ludwig-Maximilians-Universit\"at, Theresienstra{\ss}e 37, 80333 M\"unchen, Germany} 
\affiliation{Max-Planck-Institut f\"ur Physik (Werner-Heisenberg-Institut), F\"ohringer Ring 6, 80805 M\"unchen, Germany} 

\author{Alessandro Mirizzi} 
\email{alessandro.mirizzi@desy.de}
\affiliation{II Institut f\"ur Theoretische Physik, Universit\"at Hamburg, 
Luruper Chaussee 149, 22761 Hamburg, Germany} 

\author{Marco Roncadelli} 
\email{marco.roncadelli@pv.infn.it}
\affiliation{INFN, Sezione di Pavia, Via A.~Bassi 6, 27100 Pavia, Italy}


\begin{abstract}
The mixing of photons with axion-like particles (ALPs) in the large-scale magnetic field $B$ changes the polarization angle of a linearly polarized photon beam from active galactic nuclei in radio galaxies as it propagates over cosmological distances. Using available ultraviolet polarization data concerning these sources we derive a new bound on the product of the photon-ALP coupling $g_{a\gamma}$ times $B$. We find $g_{a\gamma} B \lesssim 10^{-11}$~GeV$^{-1}$~nG for ultralight ALPs with $m_a \lesssim 10^{-15}$~eV. We compare our new bound with the ones present in the literature and we  comment about possible improvements with observations of more sources. 
\end{abstract}


\preprint{LMU-ASC 16/12,  MPP-2012-61}
\maketitle

\section{Introduction}
Many extensions of the Standard Model predict the existence of axion-like particles (ALPs), namely very light spin-zero bosons $a$ characterized by a two-photon vertex $a \gamma\gamma$~\cite{Jaeckel:2010ni}. In the presence of an external electromagnetic field, this coupling gives rise to photon-ALP mixing, which leads in turn to two distinct phenomena. One is {\it photon-ALP oscillations}~\cite{sikivie,Raffelt:1987im,Anselm:1987vj}, which  is exploited by the ADMX experiment~\cite{Duffy:2006aa} to search for ALP dark matter~\cite{Arias:2012mb}, by CAST to search for solar ALPs~\cite{Arik:2008mq} and by the regeneration laser experiments~\cite{Robilliard:2007bq,Chou:2007zzc,Afanasev:2008jt,Fouche:2008jk,Ehret:2010mh}. The other phenomenon consists in the {\it change of the polarization state} of a photon beam traveling in a magnetic field. More specifically, an initially linearly polarized beam propagating in a magnetic field undergoes two different effects: \emph{(a)} it acquires an elliptical polarization, and  \emph{(b)} the ellipse's major axis gets rotated with respect to the initial polarization direction. The former effect is called \emph{birefringence} while the latter \emph{dichroism}~\cite{Maiani:1986md,Raffelt:1987im,Gasperini:1987da}. A claim for a positive observational evidence of both vacuum birefringence and vacuum dichroism by the PVLAS collaboration~\cite{Zavattini:2005tm} employing a laser beam has subsequently been withdrawn~\cite{Zavattini:2007ee}.

Over the last few years it has been realized that the $a\gamma\gamma$ coupling can also produce detectable effects in the observations of distant active galactic nuclei (AGN) and gamma-ray bursts (GRBs), since photons emitted by these sources can mix with ALPs during their propagation through large-scale magnetic fields (see~\cite{Jaeckel:2010ni} for a recent review). In this context, photon-ALP oscillations~\cite{Mirizzi:2007hr,HS:2007,HSI:2007,De Angelis:2007yu} provide a natural mechanism to drastically reduce photon absorption caused by extragalactic background light above roughly $100 \, {\rm GeV}$ where observations of AGN are currently performed by Imaging Atmospheric Cherenkov Telescopes~\cite{De Angelis:2007dy,Simet:2007sa,DMPR:2009,Mirizzi:2009aj,SanchezConde:2009wu,DeAngelis:2011id,Dominuez:2011id,Horns:2012fx,Tavecchio:2012um}. Astrophysical observations of AGN and GRBs also offer the opportunity to look for ALP effects in polarization measurements. Constraints on ALPs from future GRB polarimetric measurements have been recently discussed~\cite{Bassan:2010ya,Mena:2011xj}. 

In this paper we derive a new bound on the photon-ALP coupling using ultraviolet (UV) data from AGN in radio galaxies (RGs) by searching for a rotation of the detected photon linear polarization with respect to the emitted one. Our analysis has been inspired by recent studies~\cite{Alighieri:2010eu,Kamionkowski:2010ss}, where the same data-set has been used to constrain the effects of a possible cosmological dichroism, occurring when photons coupled to an hypothetical background of cosmic Nambu-Goldstone fields propagate over long distances. However, our investigation is quite different in spirit. For, we do not absolutely assume the presence of a background of ALPs but merely the existence of a large-scale magnetic field {\bf B}. 

Now, in spite of the fact that AGN in RGs are very complicated objects, the scattering model of anisotropic nuclear UV radiation predicts that the angle between the UV axis of the source and the UV linear polarization is $90^{\circ}$~\cite{Alighieri:2010eu}. Cosmic dichroism or photon-ALP mixing can -- by rotating the linear UV polarization -- give rise to an angle different from $90^{\circ}$ between the observed linear polarization of the UV radiation and the UV axis of a single source. Given that no rotation has been detected so far within a few degrees for each analyzed source, we are able to set a bound on the product of the photon-ALP coupling times the strength of the large-scale magnetic field -- namely on 
$g_{a\gamma} B$ -- since any observable effect involving ALPs in an external magnetic field depends on the product $g_{a\gamma} B$.

The plan of the paper is as follows. In Sec.~2 we review the formalism describing the photon-ALP mixing and its implications for photon oscillations and polarization. In Sec.~3 we introduce our input parameters, the AGN in RGs data-set and we describe the constraints coming from the absence of rotation of the linear polarization of the UV radiation emitted by the considered sources. Finally, in  Sec.~4  we compare our new bound with the the ones present in the  literature, and we discuss possible improvements.  

\section{Mixing of photons with axion-like particles}

\subsection{Oscillations}

Photon-ALP mixing occurs in the presence of an external magnetic field ${\bf B}$ due to the interaction term~\cite{Raffelt:1987im,sikivie,Anselm:1987vj}
\begin{equation}
\label{mr}
{\cal L}_{a\gamma}=-\frac{1}{4} \,\gag
F_{\mu\nu}\tilde{F}^{\mu\nu}a=\gag \, {\bf E}\cdot{\bf B}\,a~,
\end{equation}
where $\gag$ is the photon-ALP coupling constant (which has the dimension of an inverse energy). 

We consider throughout a monochromatic and linearly polarized photon beam of energy $E$ propagating along the $z$ direction in a cold ionized and magnetized medium. It has been shown that for very relativistic ALPs, the beam propagation equations takes the form~\cite{Raffelt:1987im}
\begin{equation}
\label{we} 
\left(i \, \frac{d}{d z} + E +  {\cal M} \right)  \left(\begin{array}{c}A_x (z) \\ A_y (z) \\ a (z) \end{array}\right) = 0~,
\end{equation}
where $A_x (z)$ and $A_y (z)$ are the  photon linear polarization amplitudes along the $x$ and $y$ axis, respectively, $a (z)$ denotes the ALP amplitude  and ${\cal M}$ represents the photon-ALP mixing matrix. 

Actually, ${\cal M}$ takes a simpler form if we restrict our attention to the case in which ${\bf B}$ is homogeneous. We denote by ${\bf B}_T$ the transverse magnetic field, namely its component in the plane normal to the beam direction. We can choose the $y$ axis along ${\bf B}_T$ so that $B_x$ vanishes. 
The linear photon polarization state parallel to the transverse field direction ${\bf B}_T$ is then denoted by $A_{\parallel}$ and the orthogonal one by $A_{\perp}$. In this case the mixing matrix can be written as~\cite{Mirizzi:2005ng,Mirizzi:2006zy}
\begin{equation}
{\cal M}_0 =   \left(\begin{array}{ccc}
\Delta_{ \perp}  & \Delta_R & 0 \\
\Delta_R &  \Delta_{ \parallel}  & \Delta_{a \gamma}  \\
0 & \Delta_{a \gamma} & \Delta_a 
\end{array}\right)~,
\label{eq:massgen}
\end{equation}
whose elements are~\cite{Raffelt:1987im}
$\Delta_\perp \equiv \Delta_{\rm pl} + \Delta_{\perp}^{\rm CM},$ 
$ \Delta_\parallel \equiv \Delta_{\rm pl} + \Delta_{\parallel}^{\rm CM},$
$\Delta_{a\gamma} \equiv {1}/{2} g_{a\gamma} B_T$, 
$
\Delta_a \equiv - {m_a^2}/{2E} 
$. The term
$
\Delta_{\rm pl} \equiv -{\omega^2_{\rm pl}}/{2E}$ 
where 
\begin{equation}
\omega_{\rm pl} = \left(\frac{4 \pi \alpha n_e}{m_e} \right)^{1/2} \simeq 1.17 \times 10^{- 14} \left(\frac{n_e}{{
10^{-7}\rm cm}^{- 3}} \right)^{1/2} {\rm eV}
\label{theta1}
\end{equation}
is the plasma frequency expressed as a function of the electron density in the medium $n_e$. The  terms $\Delta_{\parallel,\perp}^{\rm CM}$ describe the Cotton-Mouton effect, i.e.~the birefringence of fluids in the presence of a transverse magnetic field where $|\Delta_{\perp}^{\rm CM}- \Delta_{\parallel}^{\rm CM}| \propto B^2_T$. The latter terms are of little importance for the following analysis and will thus be neglected. The Faraday rotation term $\Delta_R$, which depends on the energy and the longitudinal component $B_z$, couples the modes $A_{\parallel}$ and $A_{\perp}$~\cite{Dasgupta:2010ck}. While Faraday rotation is in principle important when analyzing polarized photon sources, it actually plays no role at the energies we are dealing with. 

In this simplified geometry, the  component $A_{\perp}$ decouples away, and the propagation equation reduce to a 2-dimensional problem. 
Its solution follows from the diagonalization of the 2-dimensional mixing sub-matrix of ${\cal M}_0$ through a similarity transformation performed with an orthogonal matrix, parametrized by the rotation angle $\theta$ which takes the value~\cite{Raffelt:1987im}
\begin{equation}
\theta = \frac{1}{2}\textrm{arctan}\left(\frac{2 \Delta_{a \gamma}}{\Delta_{\rm pl}-\Delta_{a}}\right) \,\ .
\label{theta}
\end{equation}
In particular, the probability for a photon emitted in the state $A_{\parallel}$ to oscillate into an ALP after traveling a distance $d$ is given by~\cite{Raffelt:1987im}
\begin{eqnarray}
P_{a \gamma}  &=& {\rm sin}^2 2 \theta \  {\rm sin}^2
\left( \frac{\Delta_{\rm osc} \, d}{2} \right) \,\ \nonumber \\
&=& (\Delta_{a \gamma} d)^2 \frac{\sin^2(\Delta_{\rm osc} d/2)}{(\Delta_{\rm osc} d/2)^2} \,\ ,
\label{conv}
\end{eqnarray}
where the oscillation wave number is
\begin{equation}
\Delta_{\rm osc} = \left[(\Delta_{a} - \Delta_{\rm pl})^2 + 4 \Delta_{a \gamma}^2 \right]^{1/2} \,\ .
\end{equation}
It proves useful to define a critical energy~\cite{De Angelis:2007yu}
\begin{equation}
E_{\rm c} \equiv \frac{E \, |\Delta_a- \Delta_{\rm pl}|}{2 \, \Delta_{a \gamma}} \,\ ,
\label{ec}
\end{equation}
in terms of which the oscillation wave number can be rewritten as
\begin{equation}
\label{a17t}
{\Delta}_{\rm osc} = 2 \Delta_{a \gamma} \sqrt{1+ \left(\frac{E_{\rm c} }{E} \right)^2} \,\ .
\end{equation}
From Eqs.~(\ref{theta})-(\ref{a17t}) it follows that 
in the energy range $ E \gg E_{\rm c}$  
 the  photon-ALP mixing is maximal ($\theta \simeq \pi/4$) and the conversion probability becomes energy-independent. 
This is the so-called {\it strong-mixing regime}. Outside this regime the conversion probability turns out to be energy-dependent and vanishingly small, so that $E_{\rm c}$ acquires the meaning of a  low-energy  cut-off. As already pointed out, it is now evident that all results depend on the product $g_{a\gamma}B$, and not on $g_{a\gamma}$ and $B$ separately.

\subsection{Polarization}

Let us now focus our attention on the polarization effects, keeping in mind that for our monochromatic beam the electric field components are 
$E_{\parallel} \propto A_{\parallel}$ and $E_{\perp} \propto A_{\perp}$. We also denote by $\gamma_{\parallel}$ and $\gamma_{\perp}$ the beam photons described by $A_{\parallel}$ and $ A_{\perp}$, respectively, and by $\gamma_B$ a ``magnetic field photon''. Since the coupling in Eq. (\ref{mr}) has the form $A_{\parallel} \, B_T \, a$, it follows that the photons $\gamma_{\parallel}$ mix with the ALPs $a$ whereas the photons $\gamma_{\perp}$ do not mix at all.

Assuming the considered beam to be initially ($z = 0$) linearly polarized at an angle $\varphi_0$ (with respect to an arbitrarily chosen fixed direction), we address first the effect arising from the production of {\it real} ALPs, occurring via the $a \gamma\gamma$ vertex with one photon line representing a $\gamma_B$ photon. Since only ${\gamma}_{\parallel}$ mixes with $a$, the ${\parallel}$ mode gets depleted thereby decreasing $A_{\parallel}$ whereas $A_{\perp}$ is manifestly unchanged. As a consequence, we have a {\it rotation} of ${\bf A}$ as the beam propagates, and since it  occurs because of the selective loss of photons depending on their polarization the effect in question is called {\it dichroism}, owing to the analogous situation occurring in classical optics. Clearly after a distance $z$ the rotation angle is
\begin{equation}
\varphi (z) = \varphi_0 + \textrm{arctan}\left( \frac{A_{\parallel} (z)}{A_{\perp} (z)} \right) \,\ .
\end{equation}

Consider next what happens for the exchange of {\it virtual} ALPs, namely for the diagram in which two $a\gamma \gamma$ vertices are joined together by the $a$ line and two external photon lines are actually $\gamma_B$ photons. Evidently this diagram is nonvanishing for incoming $\gamma_{\parallel}$ photons while it vanishes for incoming $\gamma_{\perp}$ ones. Consequently, the index of refraction in the two modes ${\parallel}$ and ${\perp}$ is different, which means that the two modes $\gamma_{\parallel}$ and $\gamma_{\perp}$ travel at different speeds. Therefore, at any finite distance from the source, the beam polarization turns out to be {\it elliptical}. Even more explicitly, as the photon beam propagates its ${\bf A}$-vector changes both direction and magnitude so as to trace a spiral around the $z$ direction with elliptical sections. After each $2 \pi$ rotation, a different projected ellipse gets singled out in the plane perpendicular to the beam. This effect is called {\it birefringence} due to the similar situation taking place when a linearly polarized photon beam traverses an anisotropic medium. Still -- as long as birefringence alone is concerned -- all such ellipses have {\it parallel} major axes, which is just an elementary manifestation of the composition of two harmonic motions along orthogonal directions. 

So, when both dichroism and birefringence are at work an initially linearly polarized photon beam acquires an elliptical polarization with the ellipse's major axis rotated with respect to the initial polarization. We remark that a qualitatively similar situation occurs {\it in vacuum} according to quantum electrodynamics, where birefringence arises from Delbr\"uck scattering (photon scattering in a magnetic field)~\cite{Delbruck} while dichroism is produced by photon splitting~\cite{Adler} (see~\cite{Roncadelli} for a review).

Elsewhere the birefringence in the emission of GRBs has been analyzed~\cite{Bassan:2010ya}, whereas here our interest is focussed on dichroism. In spite of the difference in the physical effects, both studies are based on the propagation of photons from distant sources through large-scale magnetic fields. According to the standard lore, they are modeled as a network of magnetic domains, with size equal to the coherence length. In every domain the magnetic field ${\bf B}$ is assumed to have the same strength but its direction is allowed to change randomly from one domain to another. Therefore the propagation over many magnetic domains is a truly 3-dimensional problem, because -- due to the randomness of the direction of ${\bf B}$ -- the mixing matrix ${\cal M}$ entering the beam propagation equation cannot be reduced to a block-diagonal form similar to ${\cal M}_0$ in all domains. Rather, we take the $x$, $y$, $z$ coordinate system as fixed once and for all, and -- denoting by $\psi_n$ the angle between $B_T$ and the $y$ axis in the $n$-th domain -- we treat every $\psi_n$  as a random variable in the range $0 \leq \psi \leq 2 \pi$. The numerical technique adopted to solve the beam propagation equation in the present case closely follows the one used   in~\cite{Bassan:2010ya}, to which the reader is addressed for more details. We stress that because of the random orientation of the magnetic field in each domain, the effect of photon-ALP mixing on the photon polarization strongly depends on the orientation of the line of sight. As a consequence of the stochastic nature of this process, the photon polarization will be characterized by a probability distribution function, 
obtained by considering photon-ALP conversions over different realizations of the large-scale magnetic fields.

\section{AGN constraints}
\label{sec:constraints}

\subsection{Input parameters}

As a preliminary step, it is convenient to express the relevant physical quantities in units set by their natural values. The energy of UV photons is $\sim 10$~eV.
The strength of the large-scale magnetic field in the intergalactic medium has to meet the constraint $B \lesssim2.8\times10^{-7} (L/{\rm Mpc})^{-1/2}\,{\rm G}$ -- where $L$ denotes its coherence length -- which arises by scaling the original bound from the Faraday effect of distant radio sources~\cite{Kronberg:1993vk,Grasso:2000wj} to the now much better known baryon density measured by the Wilkinson Microwave Anisotropy Probe (WMAP) mission~\cite{Hinshaw:2008kr}. Its coherence length is expected to lie in the range $1 \, {\rm Mpc} < L < 10 \, {\rm Mpc}$~\cite{Blasi:1999hu}. The mean diffuse intergalactic plasma density is bounded by $n_e \lesssim 2.7 \times 10^{-7}$~cm$^{-3}$, arising from the WMAP measurement of the baryon density~\cite{Hinshaw:2008kr}. This corresponds to a plasma frequency $\omega_{\rm pl} \lesssim 1.8 \times 10^{-14}$~eV today. However, at the moment no measurements of $n_e$ are available in the large voids of the interstellar medium. Since the average value of $n_e$ inferred by WMAP has been sometimes criticized~\cite{Csaki:2001jk} claiming it to be too large for most of the intergalactic space, we allow it to vary within a range of two orders of magnitude below this value. Accordingly the upper bound on ${\omega}_{\rm pl}$ gets reduced by one order of magnitude. Recent results from the CAST experiment yield a direct bound on the photon-ALP coupling constant $\gag\lesssim 8.8\times 10^{-11}$~GeV$^{-1}$ for $m \lesssim 0.02$~eV~\cite{Arik:2008mq}, slightly better than the long-standing globular-cluster limit~\cite{Raffelt:2006cw}. In addition, for $m \lesssim 10^{- 10}$~eV a more stringent limit arises from the absence of $\gamma$-rays from SN~1987A, giving $\gag\lesssim 1\times 10^{-11}$~GeV$^{-1}$~\cite{Brockway:1996yr,Grifols:1996id} 
even if with a large uncertainty. Therefore, a suitable parametrization of the relevant quantities is
\begin{equation}
\Delta_{a\gamma} \simeq 1.52 \times 10^{-2} \left(\frac{g_{a\gamma}}{10^{-11} \, {\rm GeV}^{-1}} \right)
\left(\frac{B_T}{10^{-9} \,\rm G}\right) {\rm Mpc}^{-1}~ \,\ ,
\end{equation}
\begin{equation}
\Delta_a  \simeq -7.8 \times 10^{-3} \left(\frac{m_a}{10^{-15} \, {\rm eV}}\right)^2 \left(\frac{E}{{10 \, \rm eV}} \right)^{-1}
{\rm Mpc}^{-1}~ \,\ , 
\end{equation}
\begin{equation}
\Delta_{\rm pl}  \simeq - 1.1 \left(\frac{E}{{10\, \rm eV}}\right)^{-1} \left(\frac{n_e}{10^{-7} \, {\rm cm}^{-3}}\right) {\rm Mpc}^{-1}~ \,\ .
\end{equation}
Moreover, the value of the critical energy in Eq.~(\ref{ec}) is given by
\begin{equation}
E_{\rm c} 
\simeq  \frac{2.5 \, | m_a^2 - {\omega}_{\rm pl}^2|}{(10^{-15}{\rm eV})^2}
\left( \frac{10^{-9}{\rm G}}{B_T} \right) \left( \frac{10^{-11}\rm GeV^{-1}}{g_{a \gamma}} \right) {\rm eV}~ \,\ .
\label{eq:EL}
\end{equation}

As we are interested in studying the propagation of photons from AGN at redshifts $z \sim 3$ we must also take into account the redshift evolution of the environment that causes the photon-ALP mixing.
Notably the observed energy $E_0$ is red-shifted with respect to emitted one $E$, i.e. 
$E=E_0 (1+z)$.
 Assuming that the large-scale magnetic field is frozen in the plasma~\cite{Grasso:2000wj}, its strength scales as $B(z) = B_0 (1 + z)^2$ and clearly we have $n_e(z)=n_{e,0}(1+z)^3$. Here, the subscript 0 indicates values at the present epoch. The physical size of a magnetic domain scales as $L(z) = L_0(1 + z)^{-1}$ provided that it is smaller than the Hubble radius. 

\subsection{Data}
\label{subsec:data}

In order to quantify the amount of rotation of the linear polarization which takes place during the propagation, it is important to deal with photon sources for which the initial photon linear polarization can be reliably determined. Optical/UV radiation from AGN in RGs at redshift $z\gtrsim 2$ is very well suited in this respect. These astronomical  objects are well known emitters of linearly polarized radiation in the radio band as well as in the optical/UV bands. While the dichroism analysis of radio photons suffers from strong uncertainties, coming both from the absence of a reliable prediction for the polarization characteristics and from Faraday rotation, the analysis of optical/UV radiation is more robust. In particular, the orientation of the polarization of optical/UV radiation can be predicted from physical arguments~\cite{Cimatti:1993yc, Alighieri:2010eu} and it is expected to be {\it perpendicular} to the optical/UV source axis. Actually, astronomers define a fiducial fixed direction on the plane of the sky and define the {\it source position angle} as that formed by the optical/UV source axis with such a fiducial direction. Since photon polarization is transverse, the observed ones have polarization direction in the plane of the sky, and so one can define the {\it polarization position angle} as the one formed by the polarization and the above fiducial direction. So, we can rephrase the above expectation by stating that the difference $\Delta$P.A. between the source and the polarization position angles is expected to be $90^{\circ}$.

We consider here and show in Table~\ref{tab:sources} the set of  sources already used in~\cite{Alighieri:2010eu} to constrain the cosmological dichroism. The second column indicates the redshift $z$ of each source. The third column reports $\Delta$P.A., which has been measured on the available images. These sources were selected in order to provide robust upper limits on the rotation of the polarization angle in the UV band (1300 $\AA$ in the source rest-frame)~\cite{Alighieri:2010eu}. Indeed, the fact that $\Delta$P.A. is close to $90^\circ$ for every object -- actually compatible with $90^\circ$ within the accuracy of the measurements -- puts stringent constraints on the rotation of the linear polarization of the UV radiation during the propagation. 
\begin{table}[tdp]
\caption{Linear  UV polarization in distant AGN in RGs~\cite{Alighieri:2010eu}.}
\begin{center}
\begin{tabular}{|c|c|c|}
\hline
Name & Redshift ($z$) & $\Delta$P.A.~(\textrm{deg})          \\ \hline \hline
MRC 0211$-$122       & 2.34   & $89.0\pm3.5$       \\ \hline
4C $-$00.54                & 2.363 & $82\pm8$             \\ \hline
4C 23.56a                   & 2.482 & $94.6\pm9.7$       \\ \hline
TXS 0828+193            & 2.572 & $91.6\pm4.5$       \\ \hline
MRC 2025$-$218       & 2.63   & $86\pm9$             \\ \hline
TXS 0943$-$242        & 2.923 & $89.7\pm4.4$        \\ \hline
TXS 0119+130           & 3.516 & $95\pm16$             \\ \hline
TXS 1243+036           & 3.570 & $86.0\pm8.8$         \\ \hline 
\end{tabular}
\end{center}
\label{tab:sources}
\end{table}

\subsection{Likelihood analysis}
\label{subsec:likeli}

We derive our constraint on the rotation of the polarization angle by applying a likelihood technique according to a standard procedure~\cite{pdg}. As it is clear from the discussion presented above, this effect can depend only on two independent quantities: $g_{a\gamma}B$ and $n_{e}$, which  then span our parameter space. In order to reconstruct the probability distribution function of the ``signal'' in the presence of photon-ALP mixing we perform $10^{5}$ Monte Carlo runs of the beam propagation from the source to us in the intergalactic medium -- with the properties considered above -- for a given point in the parameter space $(g_{a\gamma}B,n_{e})$. We take $z=3$ for the redshift of the source, which corresponds to the typical redshift of the sources in Table~\ref{tab:sources}. In each simulation, the large-scale magnetic field is modeled as a network of random magnetic domains. We consider both the situation in which the magnetic field and the electron density evolve with the redshift ($z$-evolution case) and the one in which no redshift evolution takes place (no $z$-evolution case). We compute then the likelihood of the hypothesis of photon-ALP mixing by evaluating
\begin{figure}[tbp]
\begin{center}
\includegraphics[width=1.\columnwidth]{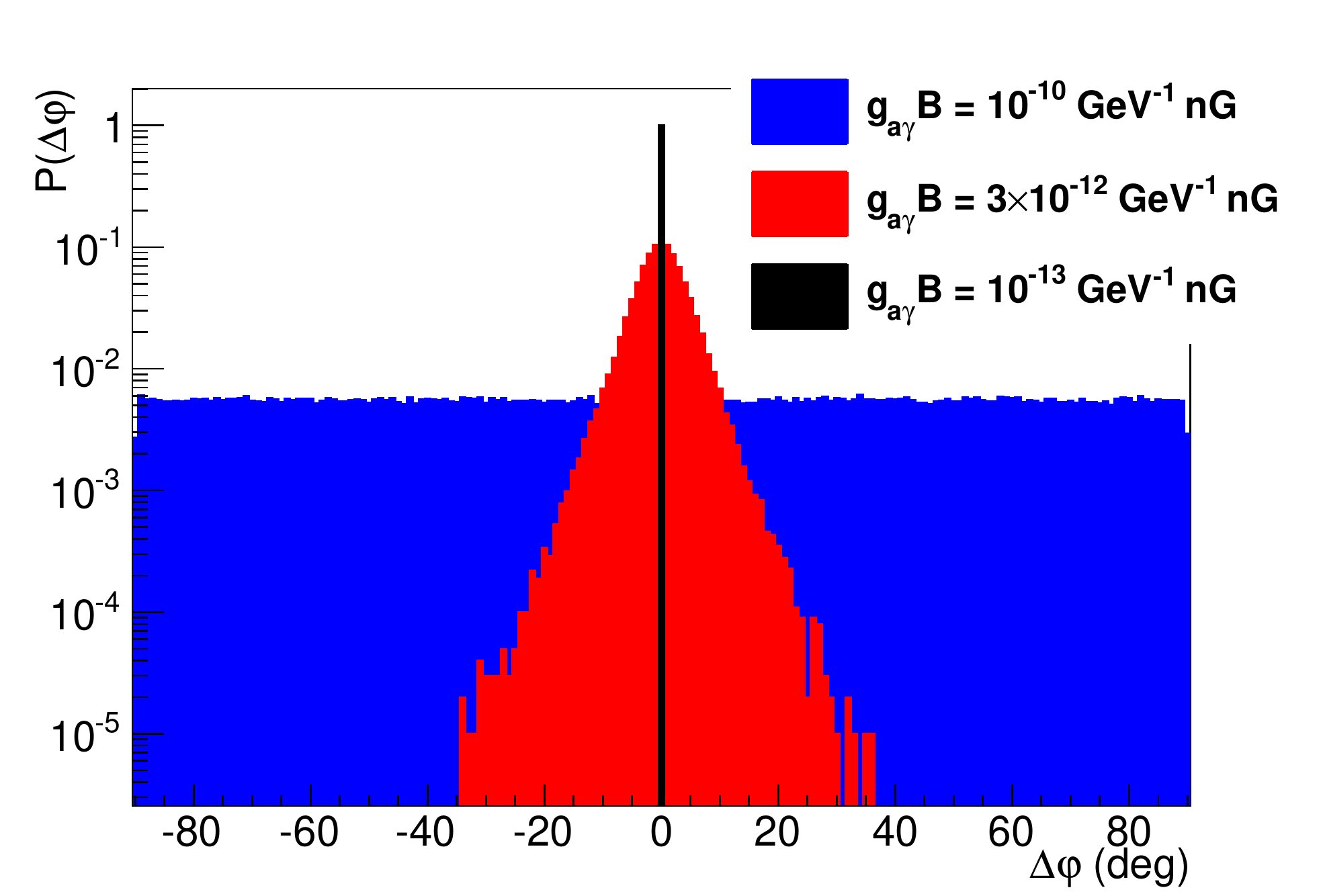}
\caption{Probability density functions $p(\Delta\varphi)$ plotted versus  the rotation of the polarization angle $\Delta\varphi$ after propagation in the extragalactic magnetic field, simulating $10^5$ AGNs at redshift $z = 3$  with initial polarization angle $\varphi_0=90^\circ$. We address the case with $z$-evolution and we take $n_e=10^{-8}$~cm$^{-3}$. Three representative values of $g_{a\gamma}B$ are considered.  }
\label{fig:figprob}
\end{center}
\end{figure}
\begin{equation}
{L}_{\rm ALP}= \Pi_{i=1}^{N} p(\varphi_{i},\sigma_{i}| \mathbf{\Theta})\;,
\label{eq:lsignal}
\end{equation}
where $\varphi_{i},~i=1...N$ denote the data on the rotation of the polarization, $\sigma_{i}$ are the errors associated with the individual measurements and 
$\mathbf{\Theta} = (g_{a\gamma}B,n_{e})$ represents the parameter space to be constrained. The function $p(x|y)$ is the probability distribution function obtained with the Monte Carlo simulations, weighted with the experimental resolution (assumed Gaussian) associated with the measurement $\varphi_{i}$
\begin{equation}
R(\varphi_{i},\sigma_{i}) = \frac{1}{\sqrt{2\pi \sigma_i^{2}}}
\exp\left(-\frac{(\varphi_i-90^{\circ})^{2}}{2\sigma_i^{2}}\right)\;.
\end{equation}
In Fig.~\ref{fig:figprob} we show the dependence of the  probability density functions $p(\Delta\varphi)$ on the rotation of the polarization angle $\Delta\varphi= \varphi(z)-\varphi_0$ with respect to the initial value $\varphi_0=90^\circ$, assuming for simplicity perfect resolution $R=1$.  As an illustration, we address the case with $z$-evolution and we take $n_e=10^{-8}$~cm$^{-3}$. In the Figure, for $g_{a\gamma} B = 10^{-13}$~GeV$^{-1}$~nG the effect on the polarization is negligible, the probability distribution being 
a delta-function peaked on $\Delta\varphi=0$. For a larger values of $g_{a\gamma} B = 3 \times 10^{-12}$~GeV$^{-1}$~nG, the mixing tends to  broaden the AGN polarization distribution around $\Delta\varphi=0$. Finally, for  $g_{a\gamma} B = 10^{-10}$~GeV$^{-1}$~nG, the probability distribution is completely flat in $\Delta\varphi$. 

The likelihood for the null hypothesis (no mixing) reads instead
\begin{equation}
{L}_{\rm null}= \Pi_{i=1}^{N} R(\varphi_{i},\sigma_{i})\;.
\end{equation}
Assuming  that Wilks' theorem applies~\cite{wilks:1938}, the confidence region compatible with the null rotation 
hypothesis is given by the condition
\begin{equation}
\textrm{ln} {{L}_{\rm ALP}}-\textrm{ln} {L}_{\rm null} \leq \frac{1}{2} \chi_\beta(k) \,\ ,
\end{equation}
where $k$ is the number of parameters, i.e. $k = 2$. Note that $\chi_\beta(k) = 2.3$, 4.61 and 6.17 for $\beta = 68.3 \%$, $90 \%$, and $95.4\%$, respectively.

\begin{figure}[tbp]
\begin{center}
\includegraphics[width=0.8\columnwidth]{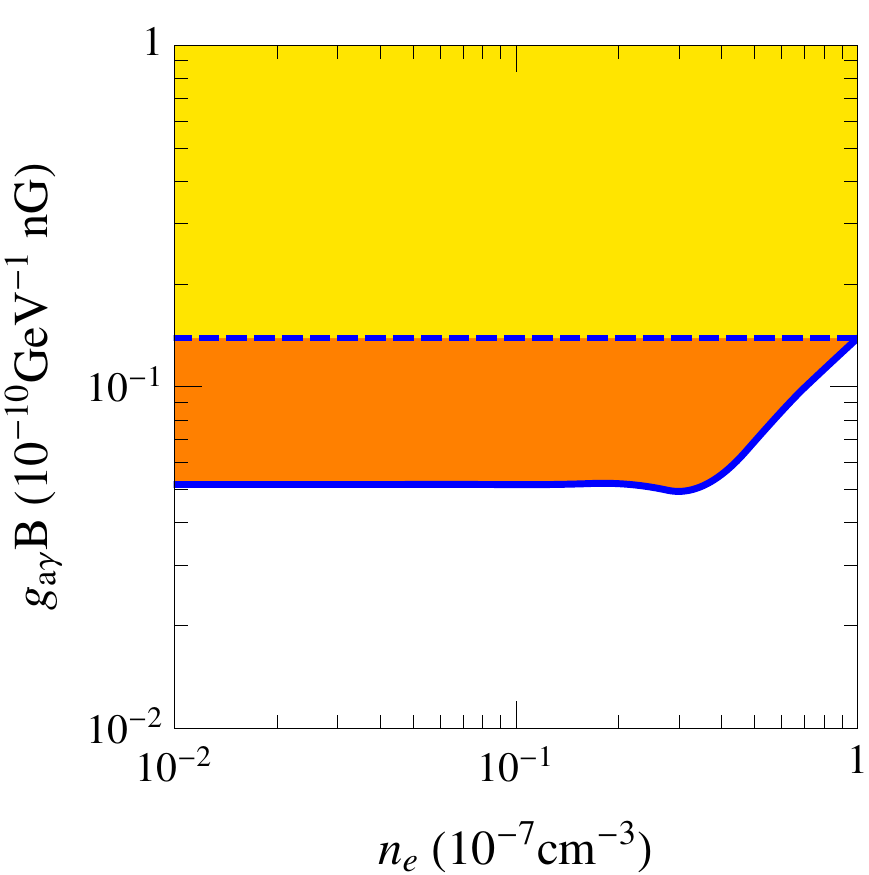}
\caption{Exclusion plot for photon-ALP conversion based on UV polarization data from AGN in RGs.
The colored regions are excluded at  95 \% CL. In particular, the region above the continuous curve is excluded for $z$-dependent case, while the region above the dashed curve panel is excluded for the $z$-independent case.}
\label{fig:conv}
\end{center}
\end{figure}

In Fig.~\ref{fig:conv} we show our exclusion contour in the plane ($n_e$, $g_{a\gamma} B$). The colored regions are excluded at 95 \% CL. In particular, 
 the region above the continuous curve is excluded in the $z$-evolution case, while the region above the dashed curve is excluded in the no $z$-evolution case. In the former case the values of $n_e$ and $g_{a\gamma}B$ shown on the axes are to be understood at $z=0$. Within a factor of a few, the same contours also hold if one increases the domain size by a factor of 10.

We find $g_{a\gamma}B \lesssim 2 \times 10^{-11}$~GeV$^{-1}$~nG in the no $z$-evolution case, and the bound improves in the $z$-evolution case 
reaching $g_{a\gamma}B \lesssim 5 \times  10^{-12}$~GeV$^{-1}$~nG for $n_e \lesssim 7 \times 10^{-8}$~cm$^{-3}$.

\section{Discussion and conclusions}
\label{sec:conclusions}

\begin{figure}[tbp]
\begin{center}
\includegraphics[width=0.8\columnwidth]{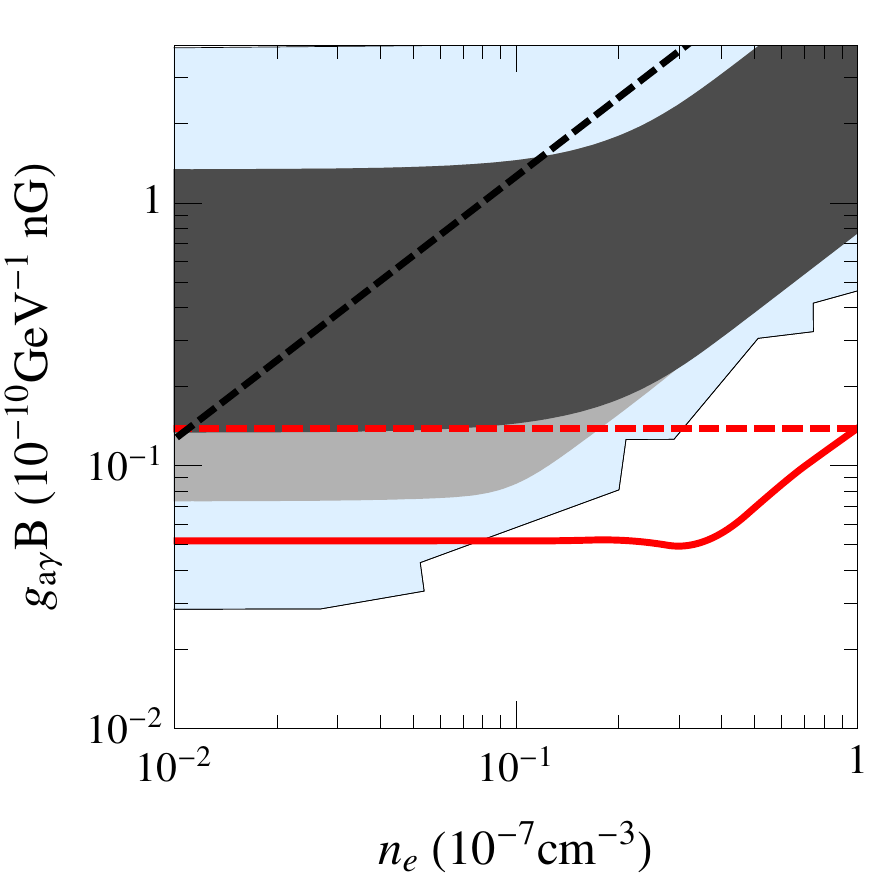}
\caption{Bounds for the product of ALP coupling to photon times magnetic field $g_{a\gamma}B$ as a function of the average electron density $n_e$. The gray
regions are excluded  from the SNIa dimming for the $z$-independent case (dark) and $z$-dependent case (light)~\cite{Avgoustidis:2010ju}. Also shown are the regions constrained by CMB~\cite{Mirizzi:2005ng} (above the black dashed line), which dominate at low $n_e$ and QSO~\cite{Ostman:2004eh} (light blue) spectral distortions. The UV exclusion region is the one above the solid curve for the $z$-dependent case, and above the dashed horizontal line for the $z$-independent case.}
\label{fig:summary}
\end{center}
\end{figure}

We have found a new limit on the quantity $g_{a\gamma} B$ for a ultralight ALPs with $m_a \lesssim 10^{-15}$~eV by studying the rotation of the linear polarization measured in UV photons from AGN in radio galaxies at cosmological distances. 

It is now interesting to compare our bound with other ones recently discussed in the literature, concerning ultralight ALPs and based on photon-ALP oscillations in the photon beam from distant sources in large-scale magnetic fields. We show in Fig.~\ref{fig:summary} a summary of these limits. 

The photon-ALP oscillations can lead to a dimming of the Type Ia supernovae luminosity curves~\cite{Csaki:2001yk,Csaki:2001jk}. This effect has been recently studied to place bounds on $g_{a\gamma} B$~\cite{Avgoustidis:2010ju}. Our limit is stronger by roughly a factor of five than the one resulting from this effect for $n_e \sim  10^{-7}$~cm$^{-3}$, while for $n_e \lesssim  10^{-8}$~cm$^{-3}$ the two bounds become comparable in the no $z$-evolution case but our one is slightly better in the $z$-evolution case.

Photon-ALP oscillations also imprint peculiar distortions on quasar (QSO) colors and spectra~\cite{Ostman:2004eh}. The corresponding bound obtained in the 
no $z$-evolution case is worse than our by a factor of three for $n_e \gtrsim  10^{-8}$~cm$^{-3}$ and becomes slightly better for lower values of the electron density.  

Finally, photon-ALP conversions can cause an excessive spectral distortion of the cosmic microwave background (CMB)~\cite{Mirizzi:2005ng,Mirizzi:2009nq}.
However, this limit would be relevant only in the case of extremely low electron density, i.e.~smaller than $\sim 10^{-10}$~cm$^{-3}$, giving $g_{a\gamma}B \lesssim \textrm{few} \times 10^{-13}$~GeV$^{-1}$~nG, or when the ALP is not ultralight and resonant photon-ALP conversion is possible~\cite{Mirizzi:2009nq}.

From our bound on $g_{a\gamma}B$ it would be possible to obtain a constraint on the photon-ALP coupling $g_{a\gamma}$. However, due to the very uncertain strength of the large-scale magnetic field $B$, this bound is uncertain by several orders of magnitude. If the large-scale magnetic field had a value  close to the upper bound $B \simeq 10~{\rm nG}$ for a domain size $L_0 \simeq 1 - 10 \, {\rm Mpc}$~\cite{Neronov:2009mj}, our astrophysical bound for ultralight ALPs would improve by at least two orders of magnitude the experimental limit from CAST~\cite{Ehret:2010mh} and by at least by one order of magnitude the one obtained from SN1987A, $g_{a\gamma} \lesssim  10^{-11}$~GeV$^{-1}$~\cite{Brockway:1996yr}. Recently, observational evidence has been discussed that the intergalactic medium is efficiently heated through generation of plasma instabilities by powerful blazars~\cite{Pfrommer:2011bg}. If this heating mechanism is at work, it would imply a model-dependent upper limit on the the large-scale magnetic field of $B \sim 10^{-3}$~nG. This would imply the  bound on the photon-ALP coupling  $g_{a\gamma} \lesssim 10^{-8}$~GeV$^{-1}$, similar to the  current one from the  ``light shining through the wall'' experiment ALPS at  DESY, namely $g_{a\gamma} \lesssim 7 \times 10^{-8}$~GeV$^{-1}$~\cite{Ehret:2010mh}.  
Conversely, if the large-scale magnetic field were close to the lower bound $B \sim 10^{-6}$~nG recently inferred from Fermi-LAT data~\cite{Ando:2010rb}, this would imply a bound on $g_{a\gamma} \lesssim 10^{-6}$~GeV$^{-1}$. It is interesting  that in this case our limit -- based on a small set of data and obtained without performing a dedicated study -- would be comparable with the bound resulting from the laser polarization experiment performed by the PVLAS collaboration~\cite{Zavattini:2007ee}.

Future improvements of our bound are expected when more data on UV polarization from distant AGN in radio galaxies will  be available. In particular, we have tested the effect of improving the precision in the determination of rotation and the effect of increasing the number of detected sources. For an error of order of $2^{\circ}$ in the determination of absence of rotation (to be compared with the $\sim6^{\circ}$ error in present data) we find only little improvement. On the other hand, increasing the number of detected sources from 8 to 20, and considering again an error of $\sim2^{\circ}$ would improve by roughly a factor of two the constraint on $\gag B$. In this respect, we remark that the number of the considered sources at $z>2$ is close to 200 \cite{Miley:2008} of which 10 objects have been observed with spectro-polarimeters and 8 have been found to be polarized. Therefore, it is not unrealistic to expect an improvement in the statistics of the UV polarized sources in question in the near future.

In conclusion, our work suggests once more the interesting physics potential of astrophysical observations as a means to constrain ALPs.  From the above discussion it follows that all the current bounds on ultralight ALPs nicely merge. Nevertheless, since every experimental measure and every cosmological and astrophysical constraint has its own systematic uncertainties and its own recognized or un-recognized loopholes, it is important to use many different approaches in order to constrain these elusive particles.

\section*{Acknowledgments} 

We thank Sperello di Serego Alighieri for many clarifying discussions and Javier Redondo for sharing with us the data we used in Fig.~2. We thank Georg Raffelt and Javier Redondo for reading this draft and for useful comments on it. The work of  A.M. was supported by the German Science Foundation (DFG) within the Collaborative Research Center 676 ``Particles, Strings and the Early Universe''. LM acknowledges support from the Alexander von Humboldt foundation.



\end{document}